\documentclass[11pt]{article}
\font\cero=cmbx10 scaled 1728 %
\font\uno=cmcsc10 scaled 1200 %
\font\dos=cmti10 scaled 1200 %
\font\tres=cmbx12 scaled 1200 %
\title{\cero Mirror potentials in classical mechanics}
\author{{\uno G.F.\ Torres del Castillo} \\
{\dos Departamento de F\'{\i}sica Matem\'atica, Instituto de Ciencias} \\
{\dos Universidad Aut\'onoma de Puebla, 72570 Puebla, Pue., M\'exico} \\
{\uno I.\ Rubalcava Garc\'{\i}a} \\
{\dos Facultad de Ciencias F\'{\i}sico Matem\'aticas} \\
{\dos Universidad Aut\'onoma de Puebla, Apartado postal 1152,} \\
{\dos 72001 Puebla, Pue., M\'exico}}
\date{ }
\begin{document}
\maketitle
\section*{ }
It is shown that for a central potential that is an injective
function of the radial coordinate, a second central potential can
be found that leads to trajectories in the configuration space and
the momentum space coinciding, respectively, with the trajectories
in the momentum space and the configuration space produced by the
original potential.\\[1ex]
{\it Keywords:} Hamiltonian mechanics; classical mechanics \\[2ex]
Se muestra que para un potencial central que sea una funci\'on
inyectiva de la coordenada radial, se puede hallar un segundo
potencial central que lleva a trayectorias en el espacio de
configuraci\'on y en el de momentos que coinciden,
respectivamente, con las trayectorias en el espacio de momentos y
de configuraci\'on producidas por el potencial original.\\[1ex]
{\it Descriptores:} Mec\'anica hamiltoniana; mec\'anica cl\'asica \\[2ex]
PACS: 45.20.Jj, 03.65.Fd

\section*{\tres 1. Introduction}
In most examples of classical mechanics, the potential energy is a
function of the coordinates only; however, such a potential
determines the orbit of the mechanical system in configuration
space and also the evolution of the momenta of the particles of
the system. For example, the central potential $V(r) = - k/r$
(which corresponds to the so-called Kepler problem) leads to
orbits in configuration space that are conics and the trajectory
in momentum space (the {\em hodograph}) is (part of) a circle
(see, {\em e.g.}, Refs.\ 1--3). Then, one may ask if there exists
a potential that leads to orbits in the configuration space that
are (part of) circles and the hodograph is a conic.

The aim of this paper is to show that, in some cases, for a given
potential one can find a second potential (which will be referred
to as the {\em mirror potential}), depending on the coordinates
only, such that the trajectories in configuration space and in
momentum space produced by the mirror potential coincide with the
trajectories in momentum space and configuration space,
respectively, corresponding to the original potential. Our
discussion will be restricted to central potentials and we shall
show that the mirror potential can be constructed whenever the
original potential is an injective function of the radial
distance.

The existence of the mirror potential is not a trivial matter. In
fact, not every system of ordinary differential equations can be
expressed in the form of the Lagrange equations (see, {\em e.g.},
Ref.\ 4 and the references cited therein). As we shall show below,
with the replacement of the original potential by the mirror
potential, it is necessary to change the time parametrization [see
Eq.\ (\ref{tiempo})]. The use of the Hamiltonian formulation
simplifies the derivation enormously.

\section*{\tres 2. Mirror potentials}
We shall consider a particle subjected to a central potential
$V(r)$; its Hamiltonian function, expressed in terms of Cartesian
coordinates, can be taken as
\begin{equation}
H = \frac{1}{2m} (p_{x}^{2} + p_{y}^{2} + p_{z}^{2}) + V \left(
\sqrt{x^{2} + y^{2} + z^{2}} \right). \label{1}
\end{equation}
(This expression for the Hamiltonian is the standard one, but
there exist many other choices, see, {\em e.g.}, Ref.\ 5.)

The equations of motion are given by the Hamilton equations
\[
\frac{{\rm d} q^{i}}{{\rm d} t} = \frac{\partial H}{\partial
p_{i}}, \hspace{5ex} \frac{{\rm d} p_{i}}{{\rm d} t} = -
\frac{\partial H}{\partial q^{i}},
\]
and, if we interchange the coordinates and momenta in Eq.\
(\ref{1}), reversing the sign of the resulting expression we
obtain a new Hamiltonian $\widetilde{H}$, which, by means of the
Hamilton equations, will lead to the trajectories in configuration
and momentum spaces defined by $H$, interchanged. In other words,
the substitution of the Hamiltonian
\begin{equation}
\widetilde{H} = - \frac{1}{2m} (x^{2} + y^{2} + z^{2}) - V \left(
\sqrt{p_{x}^{2} + p_{y}^{2} + p_{z}^{2}} \right) \label{2}
\end{equation}
into the Hamilton equations yields the same equations of motion as
$H$ but with the coordinates and momenta interchanged.

Since we are assuming that $V$ does not depend on the time, the
evolution of the state of the system in the phase space is a curve
lying on a hypersurface $\widetilde{H} = - E$, where $E$ is some
real constant (the minus sign is introduced for convenience). From
the condition $\widetilde{H} = - E$, making use of Eq.\ (\ref{2})
we then obtain,
\[
p_{x}^{2} + p_{y}^{2} + p_{z}^{2} = \left[ F \left( E -
\frac{1}{2m} (x^{2} + y^{2} + z^{2}) \right) \right]^{2},
\]
where $F$ denotes the inverse function of $V$, whose existence
requires that $V(r)$ be an injective function. The last equation
can also be written as
\begin{equation}
\frac{1}{2m} (p_{x}^{2} + p_{y}^{2} + p_{z}^{2}) - \frac{1}{2m}
\left[ F \left( E - \frac{1}{2m} (x^{2} + y^{2} + z^{2}) \right)
\right]^{2} = 0, \label{3}
\end{equation}
which is a relation of the form $h = {\rm const.}$, with
\begin{equation}
h \equiv \frac{1}{2m} (p_{x}^{2} + p_{y}^{2} + p_{z}^{2}) -
\frac{1}{2m} \left[ F \left( E - \frac{1}{2m} (x^{2} + y^{2} +
z^{2}) \right) \right]^{2} \label{4}
\end{equation}
and $h$ is now a Hamiltonian function corresponding to a central
potential
\begin{equation}
v(r) \equiv  - \frac{1}{2m} \left[ F \left( E - \frac{r^{2}}{2m}
\right) \right]^{2} \label{5}
\end{equation}
that depends parametrically on $E$.

For instance, if $V(r) = -k/r$, where $k$ is a constant, then
$F(r) = - k/r$ and, owing to Eq.\ (\ref{5}), the corresponding
mirror potential is given by
\begin{equation}
v(r) = - \frac{1}{2m} \left( \frac{2mk}{2mE - r^{2}} \right)^{2}.
\label{7}
\end{equation}
According to the discussion above, this potential leads to orbits
in configuration space that are (arcs of) circles and the orbits
in momentum space are conics. In fact, if we consider the
Hamiltonian with the mirror potential (\ref{7}) (expressed in
polar coordinates, making use of the fact that, for a central
potential, the orbit lies on a plane)
\[
h = \frac{1}{2m} \left( p_{r}^{2} + \frac{p_{\theta}^{2}}{r^{2}}
\right) - \frac{1}{2m} \left( \frac{2mk}{2mE - r^{2}} \right)^{2},
\]
taking $h = 0$ as above and using the conservation of $p_{\theta}$
we have
\[
p_{r}^{2} + \frac{L^{2}}{r^{2}} - \left( \frac{2mk}{2mE - r^{2}}
\right)^{2} = 0
\]
where $L$ is a constant. Then, the chain rule gives
\[
\frac{{\rm d} \theta}{{\rm d} r} = \frac{{\rm d} \theta/{\rm d}
t}{{\rm d} r/{\rm d} t} = \frac{L}{r^{2} p_{r}}
\]
and therefore
\[
\frac{{\rm d} \theta}{{\rm d} r} = \pm \frac{2mE - r^{2}}{r
\sqrt{(2mk/L)^{2} r^{2} - (2mE - r^{2})^{2}}}.
\]
The solution of this last equation corresponds to a circle of
radius $|mk/L|$ whose center is at a distance $\sqrt{(mk/L)^{2}
+2mE}$ from the origin.

The proof that in all cases $h$ yields the same trajectories as
$\widetilde{H}$ can be given as follows. From Eqs.\ (\ref{2}) and
(\ref{4}) one readily verifies that
\[
h = \frac{1}{2m} \left[ F \left( - \widetilde{H} -
\frac{r^{2}}{2m} \right) \right]^{2} - \frac{1}{2m} \left[ F
\left( E - \frac{r^{2}}{2m} \right) \right]^{2}
\]
so that $\widetilde{H} = -E$ is equivalent to $h = 0$, hence, on
the hypersurface $\widetilde{H} = -E$,
\[
{\rm d} h = - \frac{1}{m} F \left( E - \frac{r^{2}}{2m} \right) F'
\left( E - \frac{r^{2}}{2m} \right) {\rm d} \widetilde{H}.
\]
(The terms proportional to ${\rm d}r$ cancel as a consequence of
the condition $\widetilde{H} = -E$.) Thus, for instance,
\begin{equation}
\frac{\partial h}{\partial q^{i}} = - \frac{1}{m} F  F'
\frac{\partial \widetilde{H}}{\partial q^{i}} = \frac{1}{m} F F'
\frac{{\rm d} p_{i}}{{\rm d} t} = - \frac{{\rm d} p_{i}}{{\rm d}
\tau}, \label{6}
\end{equation}
with $F$ and $F'$ evaluated at $E - r^{2}/2m$ and we have defined
\begin{equation}
{\rm d} \tau \equiv - \frac{m}{FF'} {\rm d} t. \label{tiempo}
\end{equation}
In a similar way, one obtains $\partial h/ \partial p_{i} = {\rm
d} q^{i}/{\rm d} \tau$. That is, the trajectories generated by $h$
coincide with those generated by $\widetilde{H}$ but have a
different parametrization (see also Refs.\ 6,7).

It may be remarked that the Cartesian coordinates are not
essential in the construction of the mirror potential given above;
in fact, in the derivation of Eqs.\ (\ref{6}) only the central
character of the potential was required.

Another simple example is given by $V(r) = \frac{1}{2} kr^{2}$
(corresponding to an isotropic harmonic oscillator); in this case
$F(r) = (2r/k)^{1/2}$ and therefore the mirror potential is
\begin{equation}
v(r) = \frac{r^{2}}{2m^{2} k} - \frac{E}{mk}, \label{8}
\end{equation}
which is essentially the original potential, and this corresponds
to the fact that, for an isotropic harmonic oscillator, the
trajectories in configuration space and in the momentum space are
both ellipses. By contrast with the potential (\ref{7}), the
potential (\ref{8}) only contains the parameter $E$ in an additive
constant that has no effect in the equations of motion.
Furthermore, in this case, $\tau = -mk t + {\rm const.}$

\section*{\tres 3. Final remarks}
Apart from the possibility of extending the main result of this
paper to noncentral potentials, another natural question concerns
finding an analog of this result in quantum mechanics.

\section*{\tres Acknowledgment}
One of the authors (I.R.G.) thanks the Vicerrector\'{\i}a de
Investigaci\'on y Estudios de Posgrado of the Universidad
Aut\'onoma de Puebla for financial support through the programme
``La ciencia en tus manos.''

\section*{References}
\newcounter{ref} \begin{list}{\hspace{1.3ex}\arabic{ref}.\hfill}
{\usecounter{ref} \setlength{\leftmargin}{2em}
\setlength{\itemsep}{-.98ex}}
\item O.L.\ de Lange and R.E.\ Raab, {\it Operator Methods in Quantum
Mechanics}, (Oxford University Press, Oxford, 1991).
\item H. Goldstein, {\it Classical Mechanics}, 2nd ed., (Addison-Wesley,
Reading, Mass., 1980).
\item G.F.\ Torres del Castillo and F.\ Aceves de la Cruz, {\it Rev.\ Mex.\
F\'{\i}s.}\ {\bf 44} (1998) 546.
\item S.K.\ Soni and M.\ Kumar, {\it Europhys.\ Lett.}\ {\bf 68}
(2004) 501.
\item G.F.\ Torres del Castillo and G.\ Mendoza Torres, {\it Rev.\ Mex.\
F\'{\i}s.}\ {\bf 49} (2003) 445.
\item G.F.\ Torres del Castillo and A.\ Bernal Bautista, {\it Rev.\ Mex.\
F\'{\i}s.}\ {\bf 46} (2000) 551.
\item G.F.\ Torres del Castillo and A.\ L\'opez Ortega, {\it Rev.\ Mex.\
F\'{\i}s.}\ {\bf 45} (1999) 1.
\end{list}
\end{document}